\documentclass[a4paper,twoside,10pt]{letter}
\usepackage{graphicx,saj,multicol,subeqnarray}
\usepackage[]{rotating}

\def\papertype{Original Scientific Paper}

\def\udc{...}
\begin{document}
\baselineskip=3.1truemm
\columnsep=.5truecm
\newenvironment{lefteqnarray}{\arraycolsep=0pt\begin{eqnarray}}
{\end{eqnarray}\protect\aftergroup\ignorespaces}
\newenvironment{lefteqnarray*}{\arraycolsep=0pt\begin{eqnarray*}}
{\end{eqnarray*}\protect\aftergroup\ignorespaces}
\newenvironment{leftsubeqnarray}{\arraycolsep=0pt\begin{subeqnarray}}
{\end{subeqnarray}\protect\aftergroup\ignorespaces}
%

% Running titles

\markboth{\eightrm OPTICAL OBSERVATIONS OF IC342 GALAXY} {\eightrm M. M. VU\v{C}ETI\'{C} et al.}

{\ }

\publ

\type

{\ }

% Title

\title{OPTICAL OBSERVATIONS OF THE NEARBY GALAXY IC342 WITH NARROW BAND [S$\mathbf{II}$] AND H$\alpha$ FILTERS.
$\mathbf{I}$\footnotemark[1] } \footnotetext[1]{Based on data
collected with 2-m RCC telescope at Rozhen National Astronomical
Observatory}

% Authors

\authors{M. M. Vu\v{c}eti\'{c}$^{1}$, B. Arbutina$^{1}$, D. Uro\v{s}evi\'{c}$^{1}$, A. Dobard\v{z}i\'{c}$^{1}$,  M. Z. Pavlovi\'{c}$^{1}$,}
\authors{T. G. Pannuti$^{2}$ and N. Petrov$^{3}$}

\vskip3mm

% Address

\address{$^{1}$Department of Astronomy, Faculty of Mathematics,
University of Belgrade,\break Studentski trg 16, 11000 Belgrade,
Serbia}

\Email{mandjelic}{math.rs}

\address{$^{2}$Department of Earth and Space Sciences, Space Science Center, Morehead State University, 235 Martindale Drive,  Morehead, KY 40351, USA}

\address{$^{3}$National Astronomical Observatory Rozhen, Institute of Astronomy, Bulgarian Academy of Sciences, 72 Tsarigradsko Shosse Blvd, BG-1784 Sofia,
Bulgaria}

% Received and Accepted dates

\dates{November , 2013}{December , 2013}

% Abstract

\summary{We present observations of the portion of the nearby spiral
galaxy IC342 using narrow band [SII] and H$\alpha$ filters. These observations were carried
out in November 2011 with the 2m RCC telescope at Rozhen National
Astronomical Observatory in Bulgaria. In this paper we report coordinates,
diameters, H$\alpha$ and [SII] fluxes for 203 HII regions detected
in two fields of view in IC342 galaxy.
The number of detected HII regions is 5 times higher than previously
known in these two parts of the galaxy.}

% Keywords (see keywords.pdf file)

\keywords{ISM: HII regions -- Galaxies: individual: IC342.}

\begin{multicols}{2}
{

% Sections

\section{1. INTRODUCTION}

IC 342 is a spiral galaxy with a nearly face-on orientation
(inclination angle $\sim 20^{\circ}$ —- Tully (1988)). It
is heavily obscured by Galactic disk, and that is why it was often
avoided in optical observations. Optical observations of IC 342 are hampered by the low Galactic latitude and accompanying high extinction along the line of sight to this galaxy: for this reason, many properties of this galaxy remain poorly known. Like for many galaxies, published distance estimates to IC 342 have varied widely (from 1.8 to 8 Mpc): in this paper, we adopt a distance to this galaxy of 3.3 Mpc that was determined by Saha et al. (2002) based on Cepheid observations located in the galaxy. In
Table 1, we give basic data on this galaxy.

We aimed to conduct a census of the emission nebulae in this galaxy: these nebulae have not been well-studied in the literature except for several prominent sources. For example, a diffuse optical counterpart to the ultra-luminous X-ray source IC 342 X-1 (also known as the "Tooth Nebula") has been discussed by many authors (Roberts et al. 2003, Bauer et al. 2003, Abolmasov et al.
2007, Feng and  Kaaret 2008, Mak et al. 2011, Cseh et al. 2012).

}
\end{multicols}

\noindent
\begin{minipage}{\textwidth}
\centerline{
 {\bf Table 1.} Data for IC342 taken from NED$^{1}$.
 }
\vskip3mm \centerline{
\begin{tabular}{@{\extracolsep{-2.0mm}}c c c c c c c c @{}}
\hline
 Right ascension  & Declination & Redshift & Velocity & Distance$^{2}$ & Angular size  &  Magnitude & Gal. extinction$^{3}$\\
  $\alpha _{\mathrm{J2000}} $ & $\delta _{\mathrm{J2000}}$ & $z$ & $v$ [km s$^{-1}$] & $d$ [Mpc] & [ $'$ ]& [mag] & [mag] \\
\hline \hline
 03 46 48.5 & +68 05 47  & 0.000103 & 31 & 3.3 & $21.4 \times 20.9$ &  9.1 & 2.024 (B) \\
\hline
\end{tabular}
} \vskip2mm
$^{1}${\footnotesize \texttt{http://ned.ipac.caltech.edu/}}\\
$^{2}${\footnotesize Saha et al. (2002)}\\
$^{3}${\footnotesize Schlafly and  Finkbeiner (2011)} \ \vskip 3mm
\end{minipage}

\begin{multicols}{2}

\noindent
This particular nebula (which is associated with an ultra-luminous X-ray source) appears to be an unusual shock-powered object. Other studies of the emission nebulae in IC 342 include the work by D'Odorico et al. (1980), who conducted an optical search for supernova remnants (SNRs) in this galaxy using optical narrow band [SII] and H$\alpha$ images and found four candidates.
Hodge and Kennicutt (1983) in their atlas of HII regions in galaxies detected 666 HII regions across the entire extent of the galaxy but only the positions of the sources were given by these authors. Recently, Herrmann et al. (2008) undertook an imaging survey using narrow band [OIII] and H$\alpha$ images to identify planetary nebulae: 165 such sources were found in this galaxy.

To improve our understanding of the properties of the emission nebulae in this galaxy, we have observed IC 342 through
narrowband H$\alpha$, red continuum and [SII] filters, {in order }
to detect resident HII regions and SNRs. Observing through these filters
 allows us to {distinguish} between HII regions and SNRs
by using the criterion that SNR candidates are those objects with
[SII]/H$\alpha > 0.4$ (see e.g. Matonick and Fesen 1997). Here, we
present detection of 203 HII regions in IC342 galaxy, while the
detection of SNR candidates will be presented elsewhere.

\section{2. OBSERVATIONS AND DATA REDUCTION}

The observations were carried  out on November 27-28 2011, with the
2 m Ritchey-Chr\'{e}tien-Coud\'{e} (RCC) telescope at the National
Astronomical Observatory (NAO) Rozhen, Bulgaria ($\varphi =
41^\circ 41' 35'' ,\ \lambda = 24^\circ  44' 30'' ,\ h = 1759$ m).
The telescope was equipped with VersArray: 1300B CCD camera with
1340$\times$1300 px array, with plate scale of 0\uu 257732/px
(pixel size is 20 $\mu$m), giving the field of view $5'45''\times
5'35''$.

We observed three fields of view (FOV), which covered the west, south
and north-east part of the IC342 galaxy (Fig. 1). Centers of the field of view are:
FOV1 --  R.A.(J2000) = 03:45:45.7, Decl.(J2000) = +68:04:11.3; FOV2 -- R.A.(J2000) = 03:46:49.9,
Decl.(J2000) = +68:00:47.6; FOV3 -- R.A.(J2000) = 03:47:12.2, Decl.(J2000) = +68:08:27.7.
FOV1 and FOV2 were observed on the first night, while FOV3 was observed on the
second night, during which conditions were non-photometric, so we
present here only objects detected in FOV1 and FOV2.

The observations were performed with the narrowband [SII],
H$\alpha$ and red continuum filters. Filter characteristics are
given in Table 2. We took sets of three images through each
filter, with total exposure time of 2700s {for} each filter. Typical
seeing was 1\uu 5 -- $2.75''$. Standard star images, bias frames
and sky flat-fields were also taken.

%%Tabela sa karakteristikama filtera
\vskip.5cm

\centerline{{\bf Table 2.} Characteristics of the narrow band
filters.} \vskip5mm \centerline{\begin{tabular}{c | c c c }
Filter & $\lambda _o \ \mathrm{[\AA\mathrm ]}$ & FWHM $\mathrm{[\AA\mathrm ]}$ & $\tau _\mathrm{max}$ [\%] \\
\hline
 [SII] & 6719 & 33 & 83.3 \\
 H$\alpha$ & 6572 & 32 & 86.7 \\
 Red cont. & 6416 & 26 & 58.0 \\
\end{tabular}}

\vskip.5cm

Basic data reduction (bias substraction and flatfielding) was done
using standard procedures in  IRAF\footnote{IRAF is distributed by
the National Optical Astronomy Observatory, which is operated by
the Association of Universities for Research in Astronomy, Inc.,
under cooperative agreement with the National Science
Foundation.}. Further data reduction (image registration,
coaddition, {sky-removal}, etc.) was performed using
IRIS\footnote{Available from {\footnotesize
\texttt{http://www.astrosurf.com/buil/}}} (an astronomical images
processing software developed by Christian Buil). Three images in
each set were combined using sigma-clipping method, and then
sky-substructed (\texttt{SUBSKY}). The commands {\texttt{MAX, MIN}}
were used for cosmetic corrections (bad pixels, cosmic rays
removal). Before coaddition, each frame was multiplied with the
factor $10^{0.4\cdot\kappa X}$, where $X$ is the air mass for
a single frame and $\kappa$ is {the extinction coefficient for each
filter, in order to} remove atmospheric extinction. Atmospheric
extinction coefficients through red continuum, H$\alpha$ and [SII]
filers were measured to be 0.10, 0.08 and 0.09 mag airmass$^{-1}$
respectively. An astrometric reduction of the images was performed
by using U.S. Naval Observatory's USNO-A2.0 astrometric catalogue
(Monet et al. 1998). Each image was then flux calibrated using the observations
of the standard star Feige 34 from Massey et al. (1988).

\end{multicols}

% Figures in ps or eps format with resolution 600dpi

\centerline{\includegraphics[bb=0 0 825 825,
keepaspectratio,width=16cm]{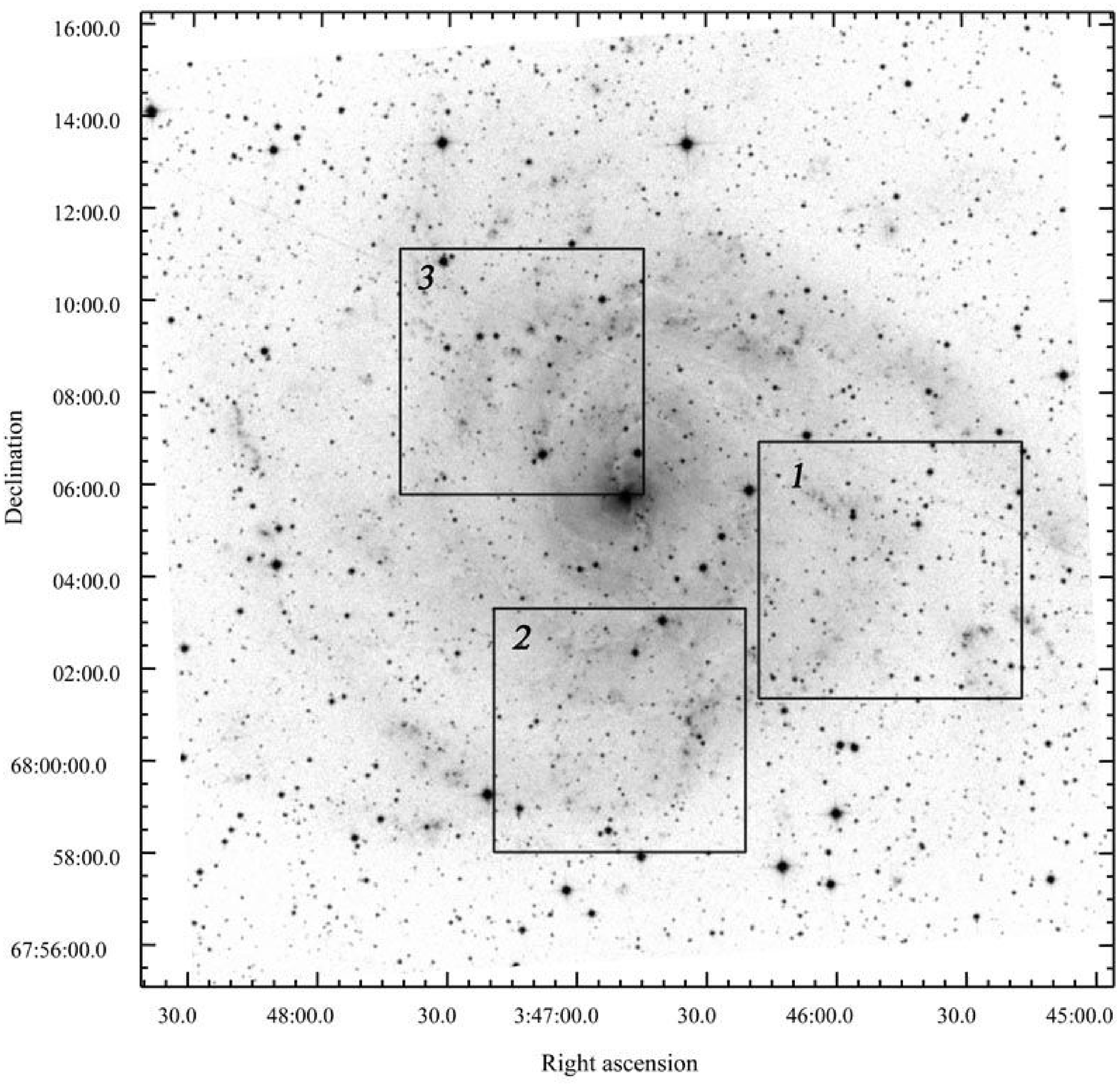}}

% Figure captions {number}{caption}

\vspace{0mm}
\begin{minipage}{14.5cm}
\centering
 \figurecaption{1.}{IC342 galaxy and observed fields of view (image from Digital Sky Survey). }
\end{minipage}

 \vspace{0mm}

\begin{multicols}{2}

Afterwards, the continuum contribution was removed from the H$\alpha$ and [SII] images.
Ratio between integrated  filter
profiles for line and continuum filters is a scaling factor
used for continuum image before it was subtracted from the
emission-line image. The next step in getting images with pure
{absolute flux-calibrated line emission} is a correction for filter
transmission.

The H$\alpha$ image ($\lambda 6563$) is contaminated with [NII]
emission ($\lambda 6548, 6583$).  To obtain the absolute
flux only from the H$\alpha$ line, we need to make corrections for
the [NII] lines as well as for the filter transmission. We use the
fact that the [NII]-6548 line is {approximately} 3 times weaker
than the [NII]-6583 line (James et al. 2005). Also, we adopt that
the integrated (sum of both components) [N II]$\lambda 6548,
6583$/H$\alpha$ ratio is 0.54 (Kennicutt et al. 2008). Knowing the
ratios between these lines and filter {transmission} at each of
them, we found that continuum-subtracted H$\alpha$ image
should be multiplied by 0.99 to get absolute flux-calibrated
H$\alpha$-line emission {(see Appendix).}

In the [SII] image, we collect emission from both [SII] $\lambda
6716$ and $\lambda 6731$ lines.  Assuming that the ratio between
them is 1.5 {(case of extremely rarefied plasma, Duric 2004)}, and
knowing filter {transmissions} at both wavelengths, we found
that the continuum-subtracted [SII] image should be multiplied
by 1.54 to obtain absolute flux-calibrated [SII]-line emission.

\end{multicols}

% Figures in ps or eps format with resolution 600dpi

\centerline{\includegraphics[bb=0 0 2700 4000,
keepaspectratio,width=14.1cm]{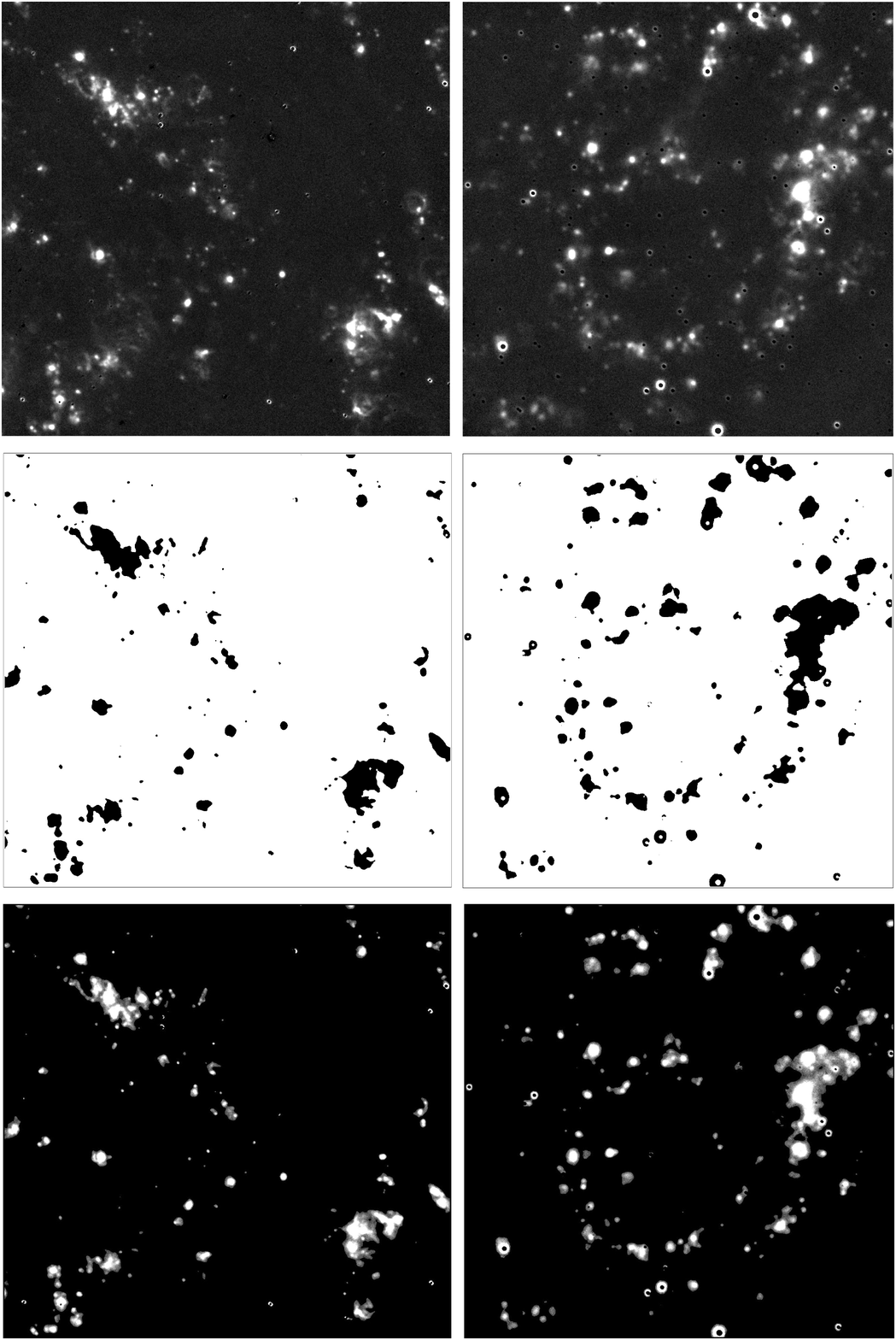}}

% Figure captions {number}{caption}

\vspace{1mm}
\begin{minipage}{14.5cm}
%\centering
 \figurecaption{2.}{Procedure for obtaining final images for photometry (FOV1 -- left, FOV2 -- right). At the top are flux-calibrated images,
 in the center are the "masks" and at the bottom are  final H$\alpha$ images.}
\end{minipage}

 \vspace{0mm}

\noindent \centerline{\includegraphics[bb=0 0 480 654,
keepaspectratio,width=\textwidth]{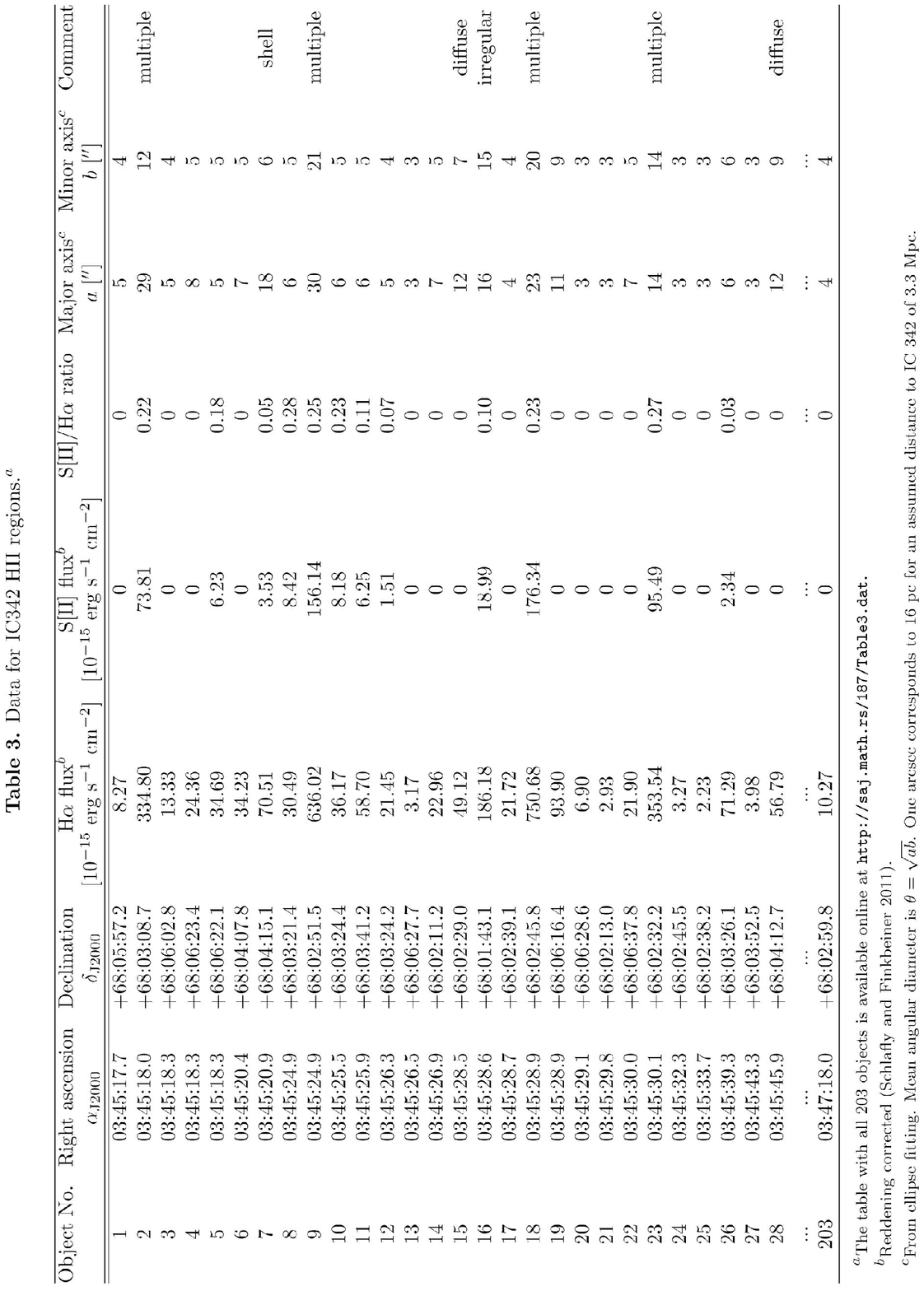}}

\clearpage

% Figures in ps or eps format with resolution 600dpi

\centerline{\includegraphics[bb=0 0 600 516,
keepaspectratio,width=16cm]{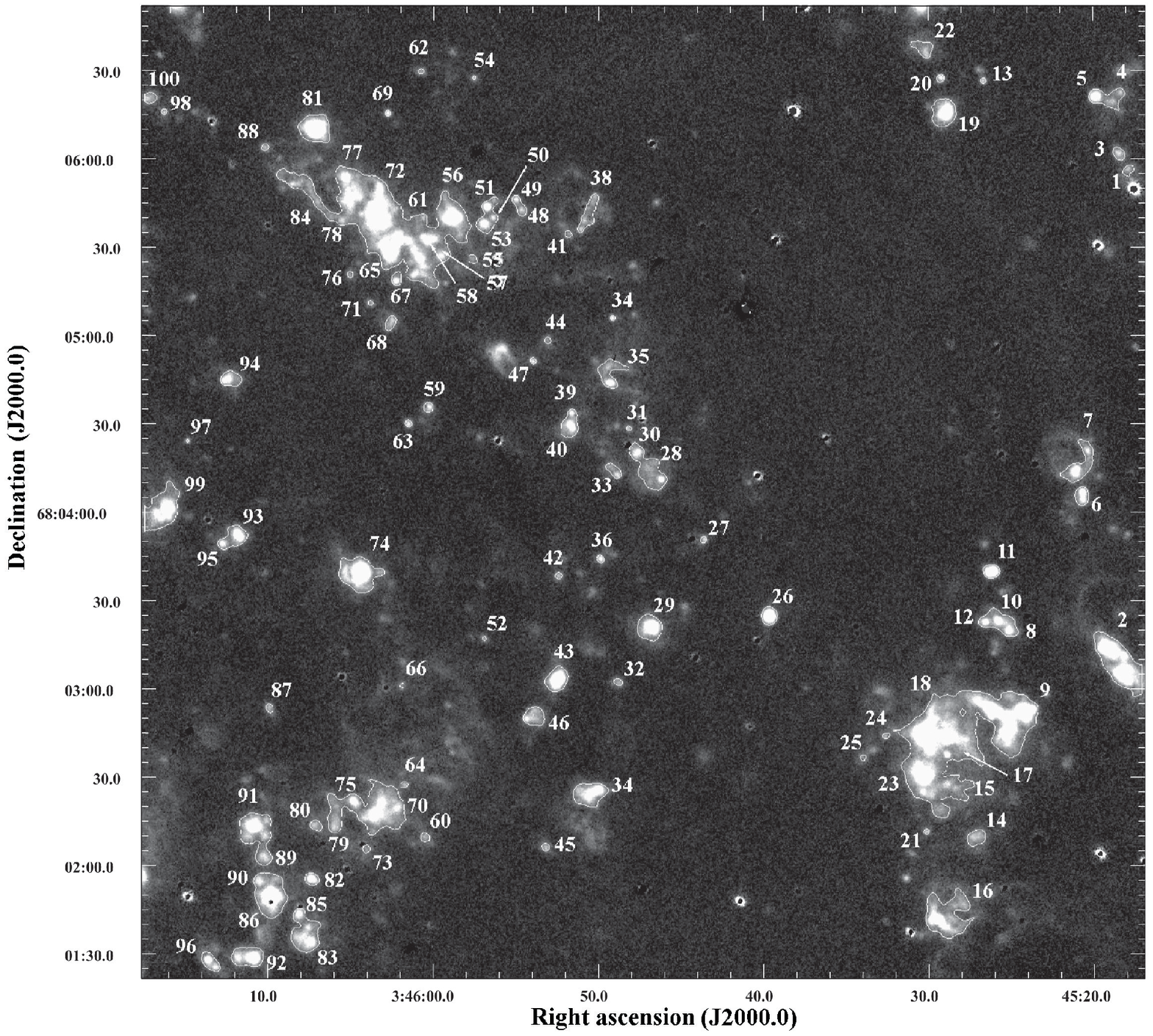}}

% Figure captions {number}{caption}

\vspace{-1mm}
\begin{minipage}{14.5cm}
\centering
 \figurecaption{3.}{The continuum-subtracted H$\alpha$ image for FOV1. Numbers correspond to the entries in Table 3.}
\end{minipage}

 \vspace{0mm}

\begin{multicols}{2}

Final images are once again background subtracted to obtain
background as flat as possible and equal to zero. After we have
removed continuum contribution, corrected H$\alpha$ emission for
the [NII]  contamination and corrected fluxes for filter
transmission, we have absolute flux-calibrated emission line
images from which we can measure fluxes of the identified objects.

\section{3. RESULTS AND CONCLUSIONS}

To extract sources from the flux-calibrated image we use the
{\texttt{BIN-UP[parameter]}} command in IRIS.  This command sets
all the pixels having an intensity higher than
{\texttt{parameter}} to 255, while other pixels are assigned value 0.
After smoothing (command {\texttt{SMEDIAN}}) {and normalizing}
this image we have the "mask". Multiplying our flux-calibrated
image with this "mask" image, gives us only sources in the image
{(Fig. 2)}. For H$\alpha$ image, we extract sources above
5$\sigma$ of the background, and from [SII] image, since
signal-to-noise ratio is lower, we extract sources above
2.5$\sigma$. Absolute fluxes are then {easily measured by} using
the single aperture, {whose} size depends on the size of the
source. Final images with the sources are given in Figs. 3 and 4.
Positions and diameters of the sources were measured by fitting
ellipse to the outer source contour, using the SAOImage DS9.

In Table 3 we give coordinates, diameters, H$\alpha$ and [SII]
fluxes, and [SII]/H$\alpha$ ratio for 203 HII regions detected in
two field of view observed in IC342 galaxy. The number of detected HII
regions is 5 times higher then the number that Hodge and Kennicutt (1983)
detected in these two parts of the galaxy.
An analysis of supernova remnant candidates revealed by our observations and identified based on their elevated [SII]/H$\alpha$ ratios will be given in a forthcoming  paper, as well as new observations planned for this galaxy.

\end{multicols}

% Figures in ps or eps format with resolution 600dpi

\centerline{\includegraphics[bb=0 0 600 516,
keepaspectratio,width=16cm]{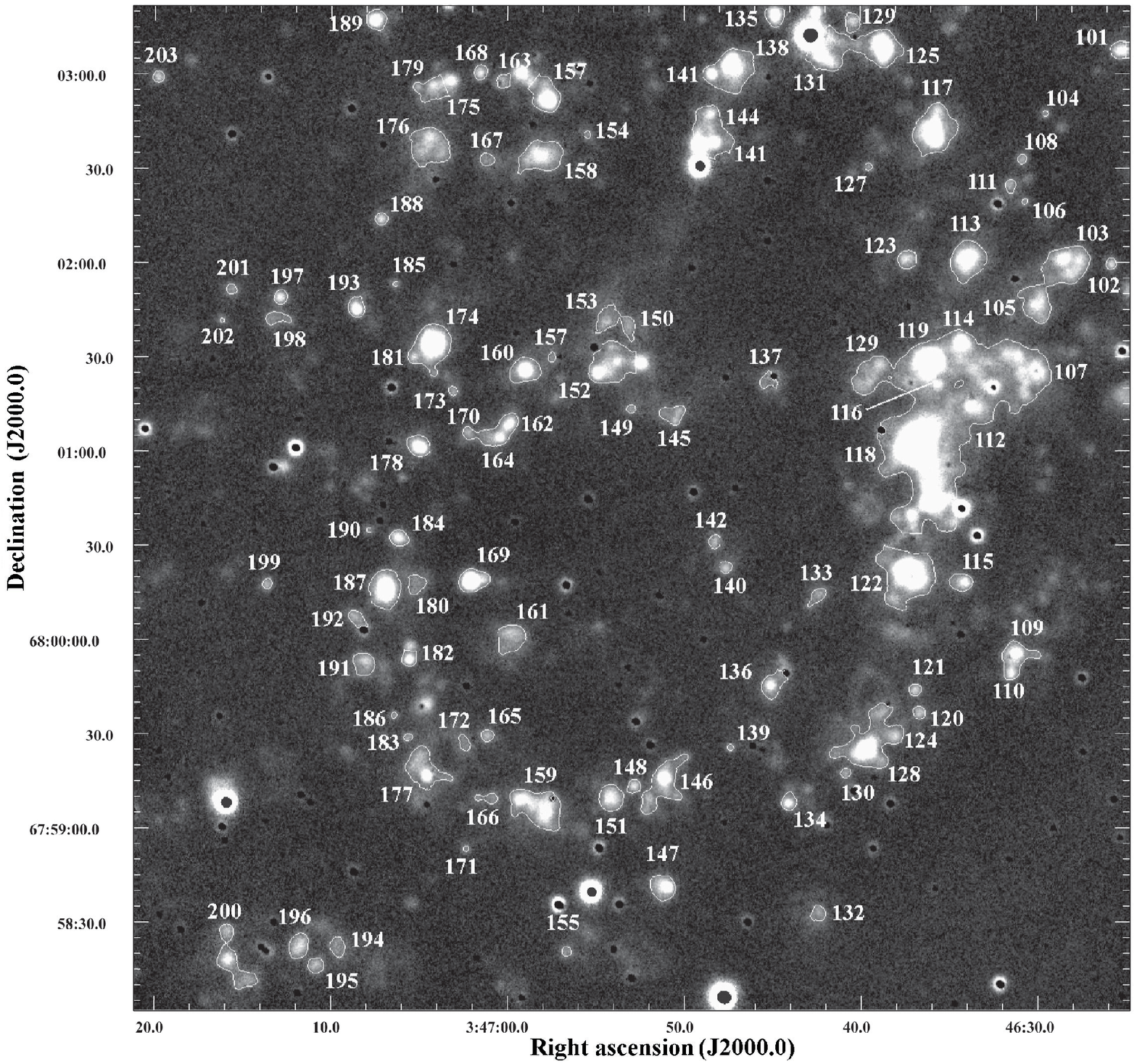}}

% Figure captions {number}{caption}

\vspace{4mm}
\begin{minipage}{14.5cm}
\centering
 \figurecaption{4.}{The continuum-subtracted H$\alpha$ image for FOV2. Numbers correspond to the entries in Table 3.}
\end{minipage}

 \vspace{3mm}

\begin{multicols}{2}

% Acknowledgements

\acknowledgements{This research has been supported by the Ministry
of Education, Science {and Technological Development} of the
Republic of Serbia through the project No. 176005 "Emission
nebulae: structure and evolution". Authors gratefully acknowledge
observing grant support from the Institute of Astronomy and Rozhen
National Astronomical Observatory, Bulgarian Academy of Sciences.}

\section{APPENDIX}

 Let us suppose that we collect emission from three lines in a
filter
\begin{equation}
I = I_0 \tau _0 + I_1 \tau _1 + I_2 \tau _2,
\end{equation}
and we want to find  $I_0$ (in our case H$\alpha$), knowing that
line ratios are $r=\frac{I_2}{I_1}$ and $R=\frac{I_1 + I_2}{I_0}$,
and filter transmissions at each of the lines are $\tau _1$, $\tau
_2$ and $\tau _3$. Then
\begin{equation}
I_0 = \frac{(1+r)I}{(1+r) \tau _0 + {R} \tau _1 + {r R} \tau _2} .
\end{equation}
If we want to find the total emission of two lines with the line
ratio $r=\frac{I_2}{I_1}$ and we measure $I = I_1 \tau _1 + I_2
\tau _2$, then
\begin{equation}
I_1 + I_2 = \frac{(1+r) I}{\tau _1 + {r} \tau _2}.
\end{equation}
which would be the  total corrected [SII] emission in our case.

% References

\references

Abolmasov P., Fabrika S., Sholukhova O. and Afanasiev V.: 2007,
\journal{Astrophysical Bulletin}, \vol{62}, 36.

Bauer, F. E.,  W. N. Brandt, W. N. and B. Lehmer, B.: 2003,
\journal{Astron. J.}, \vol{126}, 2797.

Cseh, D. et al.: 2012, \journal{Astrophys. J.}, \vol{749}, 17.

D'Odorico, S., Dopita, M. A. and Benvenuti, P.: 1980,
\journal{Astron. Astrophys. Suppl. Ser.}, \vol{40}, 67.

Duric, N.: 2004, Advanced astrophysics, Cambridge University Press, p. 270.

Feng, H. and Kaaret, P.: 2008, \journal{Astrophys. J.}, \vol{675}, 1067.

Herrmann, K. A., Ciardullo, R., Feldmeier, J. J. and Vinciguerr, M.: 2008, \journal{Astrophys. J.}, \vol{683}, 630.

Hodge, P. W. and Kennicut, R. C. Jr.: 1983, \journal{Astron. J.},
\vol{88}, 296.

James, P. A., Shane, N. S., Knapen, J. H., Etherton, J. and Percival,  S. M.: 2005, \journal{Astron. Astrophys.}, \vol{429}, 851.

Kennicutt, R. C. Jr., Lee, J. C., Funes, S. J. J. G., Sakai, S. and Akiyama, S.: 2008, \journal{ Astrophys. J. Suppl. Ser.}, \vol{178}, 247.

Mak, D. S. Y., Pun, C. S. J.  and Kong, A. K. H.: 2011, \journal{Astrophys. J.}, \vol{728}, 10.

Massey, P., Strobel, K., Barnes, J. V., and Anderson, E.: 1988, \journal{Astrophys. J.}, \vol{328}, 315.

Matonick,  D. M. and Fesen, R. A.: 1997, \journal{Astrophys. J. Suppl. Series}, \vol{112}, 49.

Monet, D. et al.: 1998, USNO-A2.0 - A catalog of astrometric
standards, U.S. Naval Observatory ({\texttt{
http://tdc-www.harvard.edu/catalogs/ ua2.html}}).

Roberts, T. P., Goad, M. R., Ward, M. J. and Warwick, R. S.: 2003, \journal{Mon. Not. R. Astron. Soc.}, \vol{342}, 709.

Saha, A., Claver, J. and Hoessel, J. G.: 2002, \journal{Astron.
J.}, \vol{124}, 839.

Schlafly, E. F. and  Finkbeiner, D. P.: 2011, \journal{Astrophys. J.}, \vol{737}, 103.

Tully, R.B.: 1988, Nearby Galaxies Catalog, Cambridge and New York, Cambridge University Press

\endreferences

\end{multicols}

\vfill\eject

{\ }

% Serbian abstract

% Title

\naslov{\noindent OPTIQKA POSMATRANJA BLISKE GALAKSIJE $\mathbf{IC342}$
KROZ USKE $\mathbf{[SII]}$ I $\mathbf{H}\alpha$ FILTERE.
$\mathbf{I}$ }

% Authors

% Authors

\authors{M. M. Vu\v{c}eti\'{c}$^{1}$, B. Arbutina$^{1}$, D. Uro\v{s}evi\'{c}$^{1}$, A. Dobard\v{z}i\'{c}$^{1}$,  M. Z. Pavlovi\'{c}$^{1}$,}
\authors{T. G. Pannuti$^{2}$ and N. Petrov$^{3}$}

\vskip3mm

% Address

\address{$^{1}$Department of Astronomy, Faculty of Mathematics,
University of Belgrade,\break Studentski trg 16, 11000 Belgrade,
Serbia}

\Email{mandjelic}{math.rs}

\address{$^{2}$Department of Earth and Space Sciences, Space Science Center, Morehead State University, Morehead, KY 40351, USA}

\address{$^{3}$National Astronomical Observatory Rozhen, Institute of Astronomy, Bulgarian Academy of Sciences, 72 Tsarigradsko Shosse Blvd, BG-1784 Sofia,
Bulgaria}

\vskip.7cm

% UDC

\centerline{UDK \udc}

% Papertype

\centerline{\rit Originalni nauchni rad}
%\centerline{\rit Prethodno saop\ss tenje}

\vskip.7cm

\begin{multicols}{2}
{

% Abstract

{\rrm U radu su predstavljena posmatranja obli\zz nje spiralne
galaksije $\mathrm{IC342}$ kroz uske $\mathrm{[SII]}$ i
$\mathrm{H}\alpha$ filtere, izvr\ss ena u novembru 2011.~godine
dvometarskim teleskopom Nacionalne astronomske opservatorije Ro\zz
en u Bugarskoj. U dva posmatrana vidna polja detektovano je ukupno 203 $\mathrm{HII}$ regiona
qiji su polo\zz aji, kao i $\mathrm{H}\alpha$ i $\mathrm{[SII]}$ fluksevi navedeni u radu.
U odnosu na prethodne studije, u ovom delu $\mathrm{IC342}$ galaksije detektovano
je 5 puta vi\ss e $\mathrm{HII}$ regiona.}

}
\end{multicols}

\end{document}